\documentclass[journal]{IEEEtran}

\usepackage{cite}
\usepackage{graphicx}
\usepackage{color}
\usepackage{amsmath}

\begin{document}

\title{Grating Lobe Suppression in Metasurface Antenna Arrays with a Waveguide Feed Layer}

\author{Michael Boyarsky,~\IEEEmembership{Student Member,~IEEE,}
        Mohammadreza F. Imani,~\IEEEmembership{Member,~IEEE,}
        \\and~David~R.~Smith,~\IEEEmembership{Senior Member,~IEEE}
\thanks{This work was supported by the Air Force Office of Scientific Research (AFOSR), grant number FA9550-18-1-0187.}
\thanks{The authors are with the Center for Metamaterials and Integrated Plasmonics, Department of Electrical and Computer Engineering, Duke University, Durham, NC, 27708. Corresponding author: michaelboyarsky@gmail.com}
\thanks{\copyright  2019 IEEE.  Personal use of this material is permitted.  Permission from IEEE must be obtained for all other uses, in any current or future media, including reprinting/republishing this material for advertising or promotional purposes, creating new collective works, for resale or redistribution to servers or lists, or reuse of any copyrighted component of this work in other works.}}

\markboth{IEEE Transactions of Antennas and Propagation}%
{Shell \MakeLowercase{\textit{et al.}}: Bare Demo of IEEEtran.cls for IEEE Journals}

\maketitle

\begin{abstract}

Metasurface antenna arrays, formed by tiling multiple metasurface subapertures, offer an alternative architecture to electrically large beamsteering arrays often used in radar and communications. The advantages offered by metasurfaces are enabled by the use of passive, tunable radiating elements. While these metamaterial elements do not exhibit the full range of phase tuning as can be obtained with phase shifters, they can be engineered to provide a similar level of performance with much lower power requirements and circuit complexity. Due to the limited phase and magnitude control, however, larger metasurface apertures can be susceptible to strong grating lobes which result from an unwanted periodic magnitude response that accompanies an ideal phase pattern. In this work, we combine antenna theory with analytical modeling of metamaterial elements to mathematically reveal the source of such grating lobes. To circumvent this problem, we introduce a compensatory waveguide feed layer designed to suppress grating lobes in metasurface antennas. The waveguide feed layer helps metasurface antennas approach the performance of phased arrays from an improved hardware platform, poising metasurface antennas to impact a variety of beamforming applications.

\end{abstract}


\IEEEpeerreviewmaketitle

\section{Introduction}
\label{s:intro}

\IEEEPARstart{M}{etasurface} antennas are an emerging alternative to conventional electronically scanned antennas that can match the performance of existing hardware with significantly improved cost and manufacturability \cite{boyarsky2017synthetic,sleasman2017experimental,watts2017x,smith2017analysis,boyarsky2018single}, with applications including security screening \cite{hunt2013metamaterial,gollub2017large} and satellite communications \cite{johnson2015sidelobe,stevenson2016metamaterial}. In this work, we examine the potential to create large metasurface antenna arrays that can be used for imaging and communication from airborne and spaceborne vehicles. With an increased demand in multi-antenna systems on mobile platforms, the hardware advantages in weight and power consumption availed by metasurface antennas are becoming increasing sought-after \cite{arapoglou2011mimo,sun2014mimo,krieger2014mimo,kim2015spaceborne,ma2018maritime,an2018topology}.

Waveguide-fed metasurface antennas use a waveguide to excite an array of metamaterial radiators. As the guided wave traverses the waveguide, each metamaterial element couples energy from the guided wave into free space as radiation. The radiation pattern of the aperture is then the superposition of the radiation from each of the element \cite{sleasman2016design}. Introducing individually addressable tunable components within each metamaterial element grants electronic control over the radiation pattern, with steered, directive beams among the available waveforms \cite{sleasman2016waveguide}. For applications requiring large reconfigurable antennas, 2D metasurface antenna arrays can be created by tiling several 1D waveguide-fed metasurfaces.

\begin{figure}
	\centering
	\includegraphics[width=8cm]{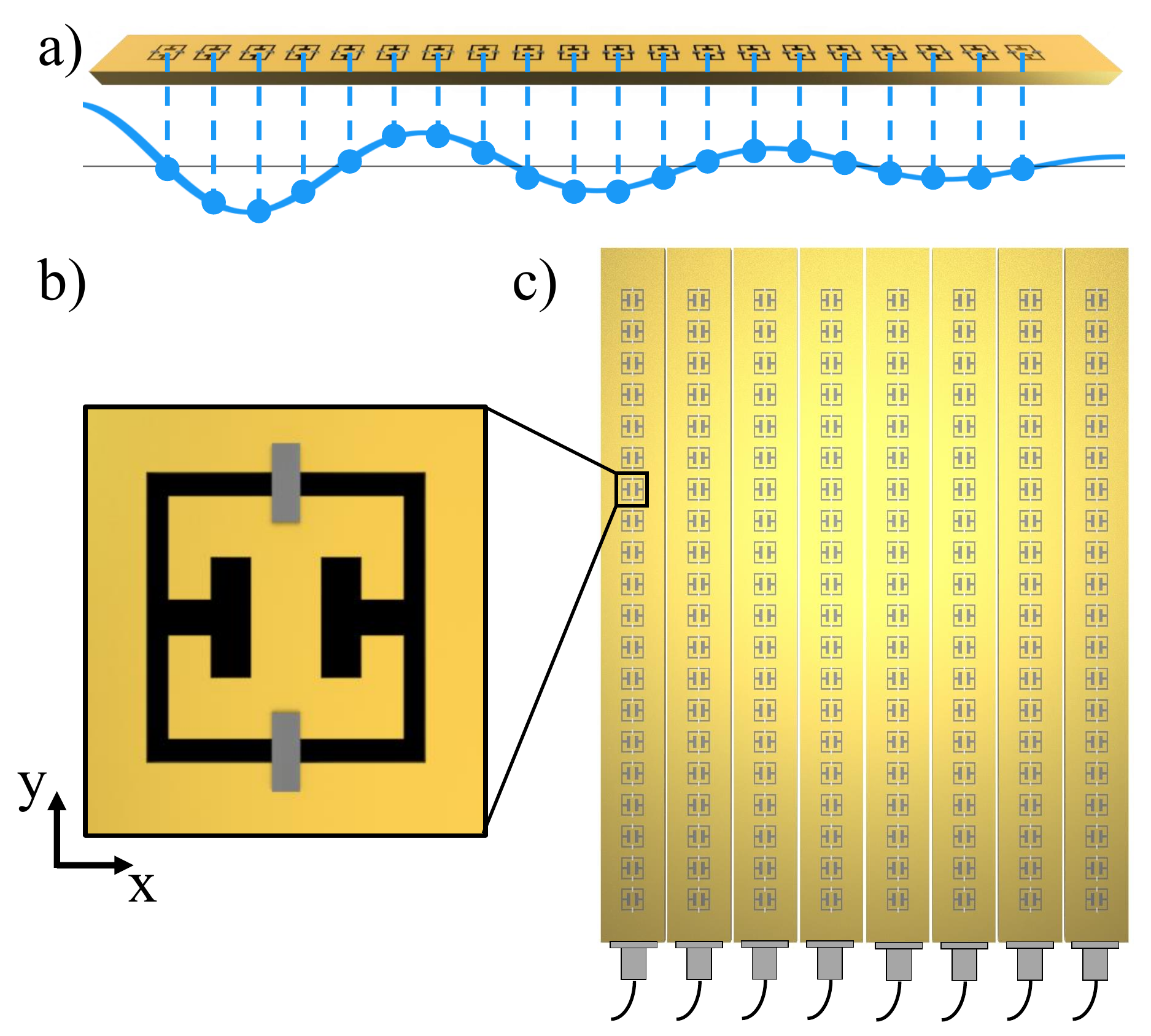}
	\caption{Diagram of an example metasurface antenna architecture. a) shows the operation of a holographic metasurface antenna, with the reference wave (shown in blue) interfering with the metamaterial elements which form the hologram \cite{boyarsky2017synthetic}. b) shows a sample metamaterial element and the surrounding conductor (shown in gold), along with the tunable component (shown in grey), which could be a varactor diode, for example. c) shows a sample antenna layout consisting of eight individually fed rectangular waveguides, each with twenty radiating elements.}
	\label{f:diagram}
\end{figure}

Metasurface antennas derive several of their advantages by exchanging tuning range in favor of low-cost, passive tuning components \cite{smith2017analysis}. Without active phase shifters and amplifiers common to conventional beamsteering devices, a metasurface antenna must be tuned by shifting the resonance of each metamaterial element. Tuning metamaterial elements this way forgoes full control over the complex response, limiting the available phase states to $-180^\circ<\phi<0$ and coupling the magnitude and phase response \cite{smith2017analysis}. These constraints can lead to coarse effective element spacing due to a periodic magnitude profile, which causes grating lobes. If each waveguide is excited with the same phase, grating lobes from each waveguide constructively interfere, magnifying their impact. In previous metasurface apertures, grating lobes were suppressed by using high dielectrics (to decrease the wavelength of the guided wave) along with dense element spacing. However, this approach can introduce practical challenges in terms of element size and efficiency, especially in the context of airborne and spaceborne systems, where antenna efficiency and structural rigidity are of utmost importance, meaning that air-filled (or vacuum-filled) waveguides must be used in favor of dielectric-filled waveguides.

In this work, we analytically examine the radiation behavior of a 2D metasurface antenna array formed by tiling 1D waveguid-fed metasurfaces. First, we describe how to tune metamaterial elements for beamforming, then we elucidate the source of the grating lobes. We then mathematically describe how varying the incident phase exciting each waveguide can suppress the grating lobes. Last, we propose and demonstrate how a compensatory, designed waveguide feed layer can optimally suppress grating lobes in metasurface antenna arrays. Using this approach, large, electronically reconfigurable, and single-port metasurface antennas can be constructed with desirable performance characteristics for applications ranging from satellite communications to earth observation.

\section{Metasurface Antenna Architecture}
\label{s:metasurface}

The waveguide-fed metasurface illustrated in Fig.~\ref{f:diagram}a consists of an array of metamaterial irises patterned into one side of a waveguide \cite{boyarsky2017synthetic,smith2017analysis}. The metamaterial elements couple energy from the waveguide mode to free space as radiation \cite{hunt2013metamaterial}. The overall radiation pattern is then the superposition of each element's radiation, whose complex amplitude is a function of the element's response and the input frequency. By altering each metamaterial element's properties, or by changing the driving frequency, different radiation patterns can be generated from the aperture. To realize an electrically large 2D antenna, 1D waveguide-fed metasurface apertures can be patterned side-by-side to form a 2D metasurface antenna array as in Fig.~\ref{f:diagram}c.

To analyze a 2D metasurface antenna analytically, we first model the constituent metamaterial elements. In this work, we assume that the elements are weakly coupled, such that inter-element coupling from the elements within the waveguide can be ignored. Other modeling methods can account for these effects, but are not used in this work \cite{pulido2017polarizability}. The simplified model used here facilitates an array factor analysis which provides sufficient insight into the grating lobe problem associated with metasurface antennas. 

Each metamaterial element can be modeled as a point dipole with a response dictated by the incident magnetic field and the element's complex polarizability (while polarizability is a tensor quantity, we assume that our elements are only polarizable in the x direction and thus use a scalar approximation) \cite{lipworth2013metamaterial,pulido2017polarizability}.

\begin{equation}
\label{eq:dipole_moment}
\eta_n=H_n\alpha_n
\end{equation}

\begin{figure}
	\centering
	\includegraphics[width=8.8cm]{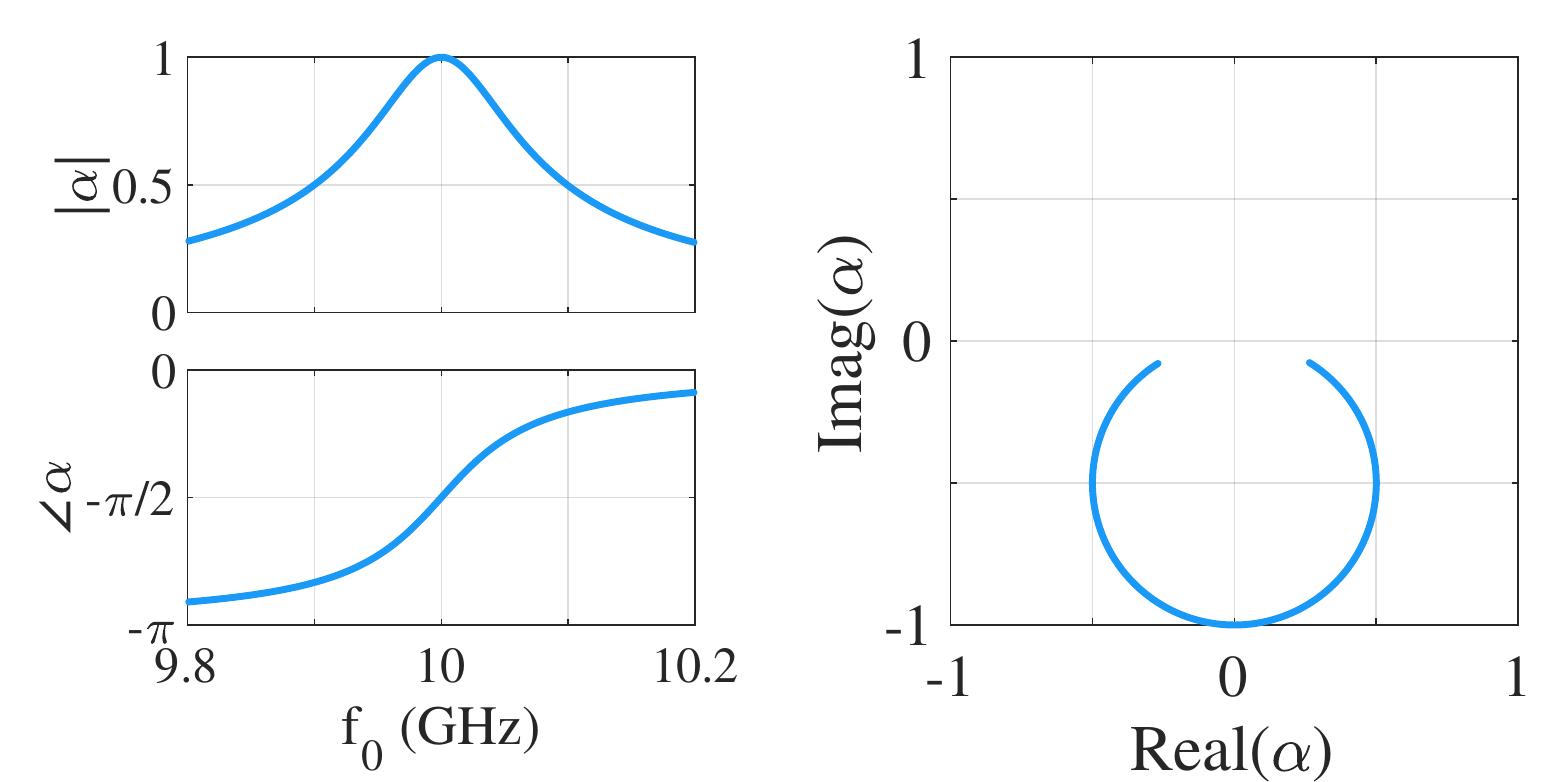}
	\caption{A sample metamaterial element's response, modeled with an analytic polarizability based on a Lorentzian response described in Eq.~\ref{eq:lorentzian}.}
	\label{f:element}
\end{figure}

\noindent Here, $\eta_n$ is the dipole moment of the $n^{th}$ element, $\alpha_n$ is the polarizability, and $H_n$ is the magnetic field, described by

\begin{equation}
\label{eq:h_field}
H_n=H_0e^{-j\beta y_n}
\end{equation}

\noindent where $\beta$ is the waveguide constant and $y_n$ is the position measured from the origin, as shown in Fig.~\ref{f:diagram}. To keep our analysis general, we model each of the metamaterial elements as having an analytic, Lorentzian polarizability \cite{smith2017analysis}

\begin{equation}
\label{eq:lorentzian}
\alpha = \frac{F\omega^2}{\omega_0^2-\omega^2+j\omega\Gamma}
\end{equation}

\noindent where $F$ is the oscillator strength ($F=1$), $\omega$ is the angular frequency, $\omega_0$ is the resonant frequency, and $\Gamma$ is the loss term ($\Gamma=7.2\times10^8$ rad/s). A sample Lorentzian element response is shown in Fig.~\ref{f:element}. Tuning an element is often accomplished by changing $\omega_0$, leading to shifts in the magnitude and phase of the polarizability. Since the polarizability is a complex analytic function of the resonant frequency, the magnitude and phase are inextricably linked, as can be expressed by the relationship \cite{smith2017analysis}

\begin{equation}
\label{eq:lorentzian2}
|\alpha|=\frac{F\omega}{\Gamma}|\cos{\psi}|,
\end{equation}

\noindent where $\psi$ denotes the phase advance introduced by the element. 

As expressed in Eq.~\ref{eq:lorentzian2} and depicted in Fig.~\ref{f:element}, the Lorentzian response has a range of phase advance restricted to $-180^\circ<\phi<0$ ---less than half of the control range of active phase shifters. Moreover, the magnitude varies over this phase range, falling to zero at either extreme such that the actual useful phase range of the element is even smaller than $180^\circ$. Using a subwavelength spacing of elements ($\lambda/4$ or denser, where $\lambda$ is the free space wavelength) and leveraging the phase accumulation of the guided wave can help to regain some of the reduced element control, but these approaches can be costly and challenging, especially at higher frequencies or with large apertures \cite{sievenpiper2003two,lim2004metamaterial,oliner2007leaky,martinez2013holographic}. When these approaches are not possible, however, the strong magnitude modulation of the element weights (which accompany a typical desired phase distribution for beamforming will produce grating lobes. For 1D antennas there is no simple way to avoid these grating lobes; for 2D arrays formed by tiling a set of 1D antennas, the additional degrees of freedom can be used to cancel out the grating lobes, providing alternative design approaches for larger waveguide-fed metasurface arrays.

To illustrate the grating lobe problem associated with metasurface antennas, we compute radiation patterns using array factor calculations, with the dipole moments from Eq.~\ref{eq:dipole_moment} as the antenna weights \cite{hansen2009phased,balanis2016antenna}

\begin{equation}
\label{eq:array_factor}
AF(\theta,\phi)=\sum_{n=1}^{N}\sum_{m=1}^{M}\eta_{n,m}e^{-jk(x_m\sin{\theta}\sin{\phi}+y_n\sin{\theta}\cos{\phi})}
\end{equation}

\noindent where $k$ is the free space wavenumber, $N$ is the number of elements (in the y direction), $M$ is the number of waveguides (in the x direction), $\theta$ is the elevation angle, and $\phi$ is the azimuth angle. Ideally, the polarizability of each element could be set to counteract the phase of the guided wave and form a beam steered to $(\theta_s,\phi_s)$, following

\begin{equation}
\label{eq:ideal_alpha}
\alpha_n=e^{j(\beta y_n +k x_m \sin\theta_s \sin\phi_s + k y_n \sin\theta_s\cos\phi_s)}.
\end{equation}

\noindent However, due to the Lorentzian-constrained nature of metamaterial elements, the polarizability profile given by Eq.~\ref{eq:ideal_alpha} cannot be fully realized. In \cite{smith2017analysis}, Lorentzian-constrained modulation (LCM) was derived, which optimizes the phase and magnitude of each metamaterial element simultaneously in order to approximate Eq.~\ref{eq:ideal_alpha}.

\begin{equation}
\label{eq:alpha_lcm}
\alpha_n=\frac{j-e^{j(\beta y_n +k x_m \sin\theta_s \sin\phi_s + k y_n \sin\theta_s\cos\phi_s)}}{2}
\end{equation}

\noindent Inserting Eq.~\ref{eq:alpha_lcm} into Eq.~\ref{eq:array_factor}, the array factor becomes

\scriptsize
\begin{equation}
\label{eq:lcm_farfield}
\begin{split}
&AF(\theta,\phi)=\frac{H_0}{2}\Bigg[\sum_{n=1}^{N}\sum_{m=1}^{M}je^{-j(\beta y_n + k y_n \sin\theta\cos\phi + k x_m \sin\theta\sin\phi)}\\
-&\sum_{n=1}^{N}\sum_{m=1}^{M}e^{-j k \big(y_n (\sin\theta \cos\phi-\sin\theta_s \cos\phi_s)+x_m (\sin\theta \sin\phi-\sin\theta_s \sin\phi_s)\big)}\Bigg]\\
\end{split}
\end{equation}
\normalsize

\begin{figure}
	\centering
	\includegraphics[width=8.8cm]{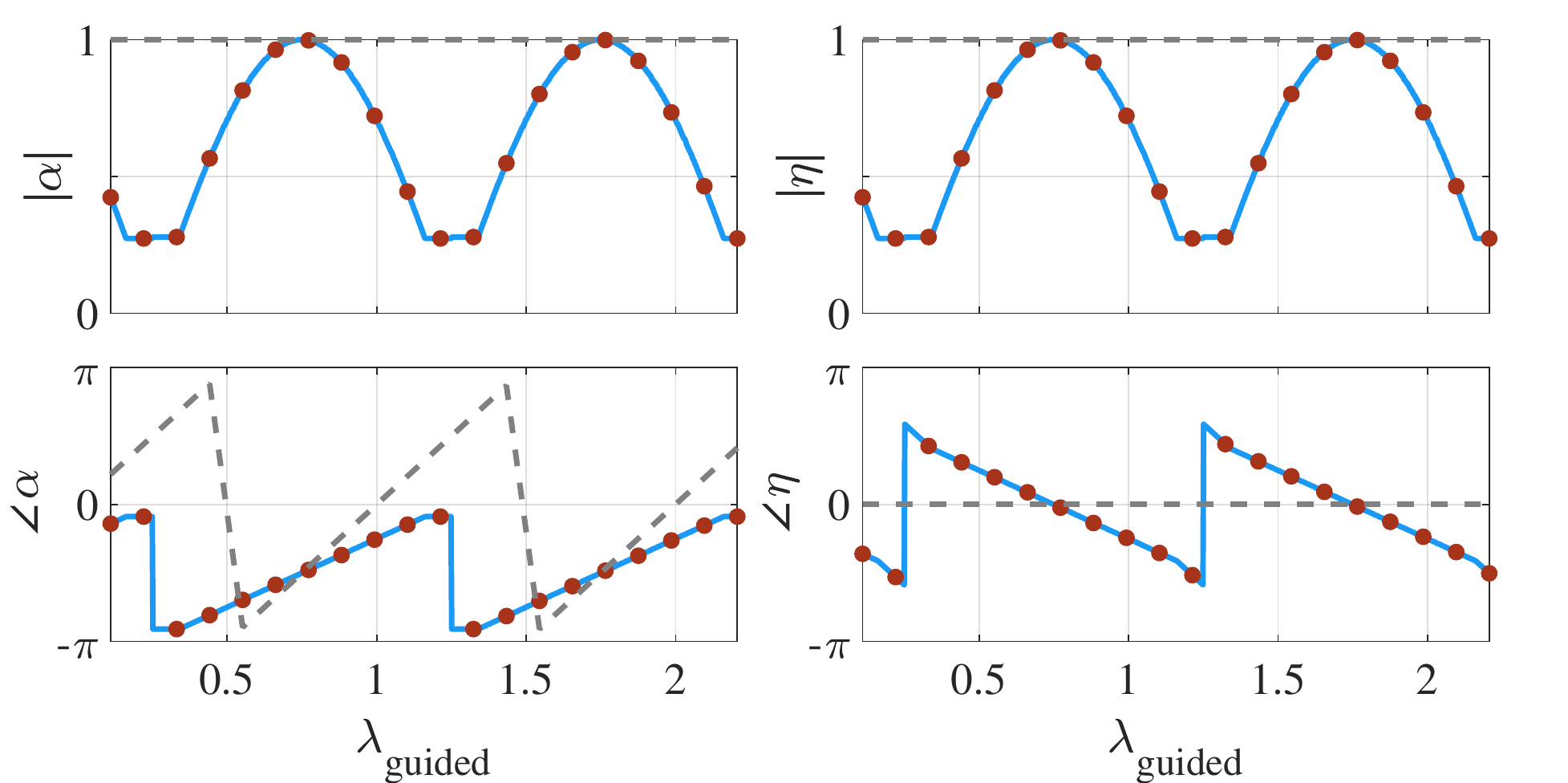}
	\caption{Mapping from desired to available $\alpha$ and corresponding $\eta$. The gray, dashed line shows the desired response, while the blue, solid line shows the achievable response. The red dots indicate the discrete locations of the metamaterial elements.}
	\label{f:mapping}
\end{figure}

\begin{figure}
	\centering
	\includegraphics[width=8.8cm]{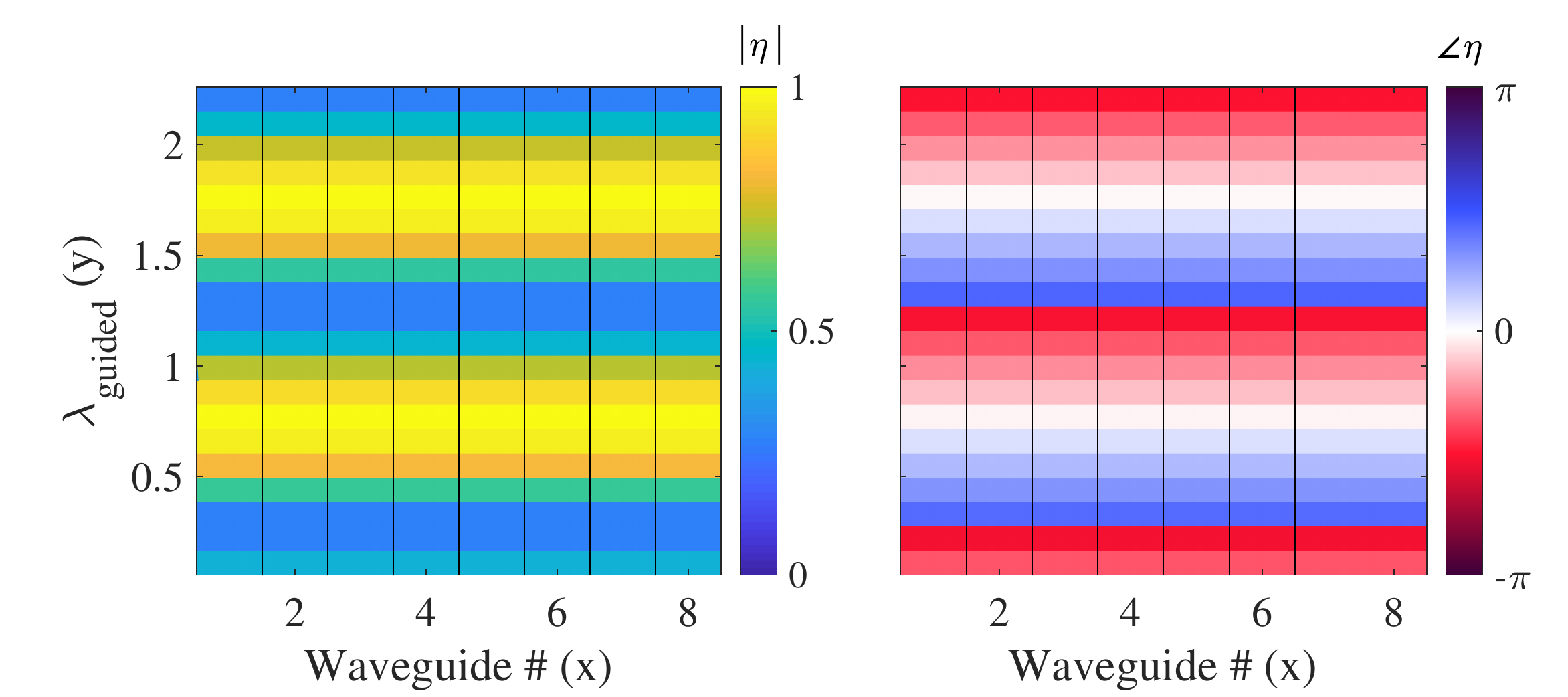}
	\caption{Antenna weights for the sample metasurface antenna shown in Fig.~\ref{f:diagram}. In the case of metamaterial radiators, the weights are given by the dipole moments, $\eta$, for which the magnitude and phase are shown.}
	\label{f:weights_no_feed}
\end{figure}

\begin{figure}
	\centering
	\includegraphics[width=7.5cm]{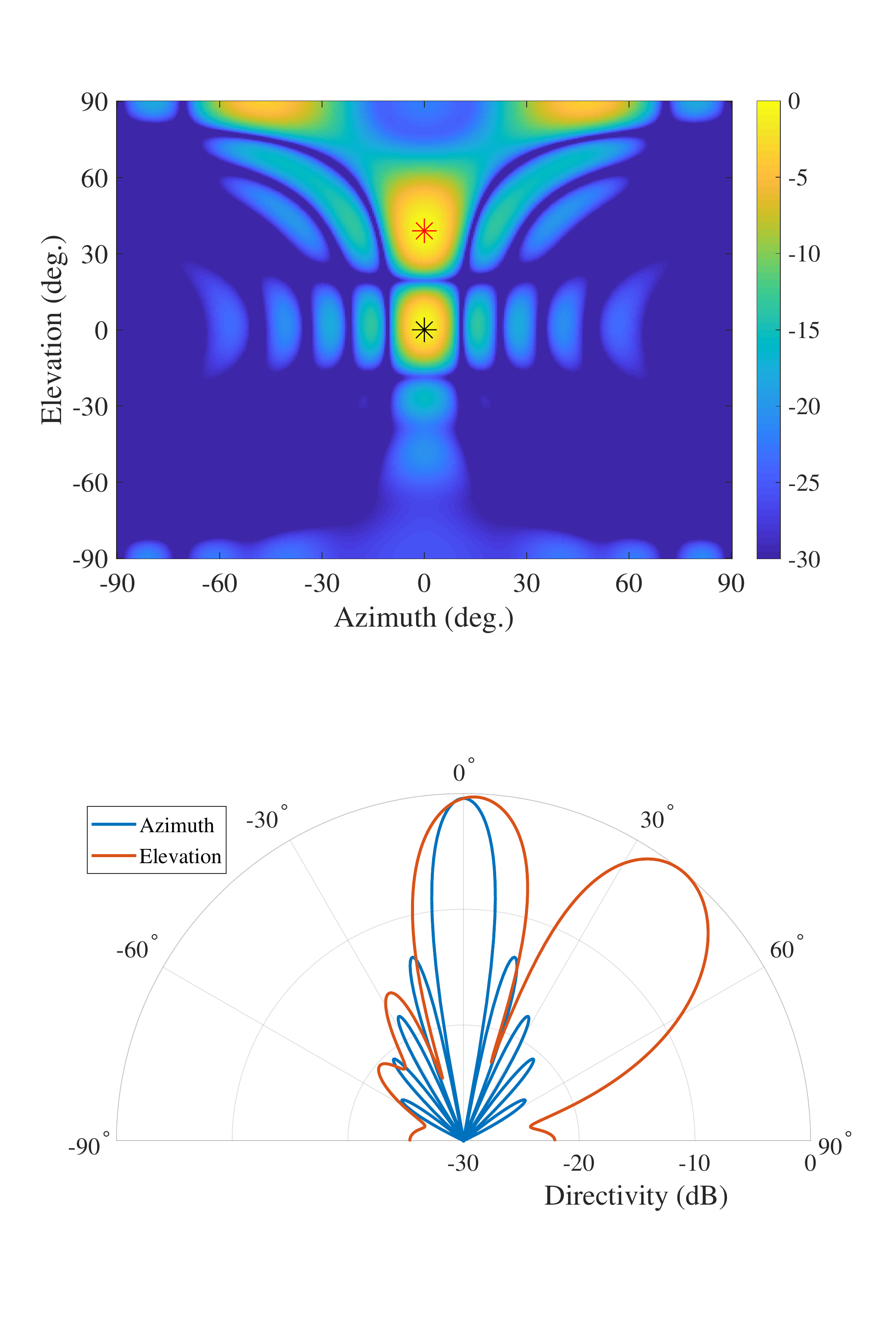}
	\caption{Normalized farfield pattern resulting from the simulated metasurface antenna array, which follows the mapping shown in Fig.~\ref{f:mapping} and \ref{f:weights_no_feed}. The beam is generated towards broadside, with the target location indicated by the black star and the grating lobe indicated by the red star.}
	\label{f:farfield_no_feed}
\end{figure}

\begin{figure}
	\centering
	\includegraphics[width=7.5cm]{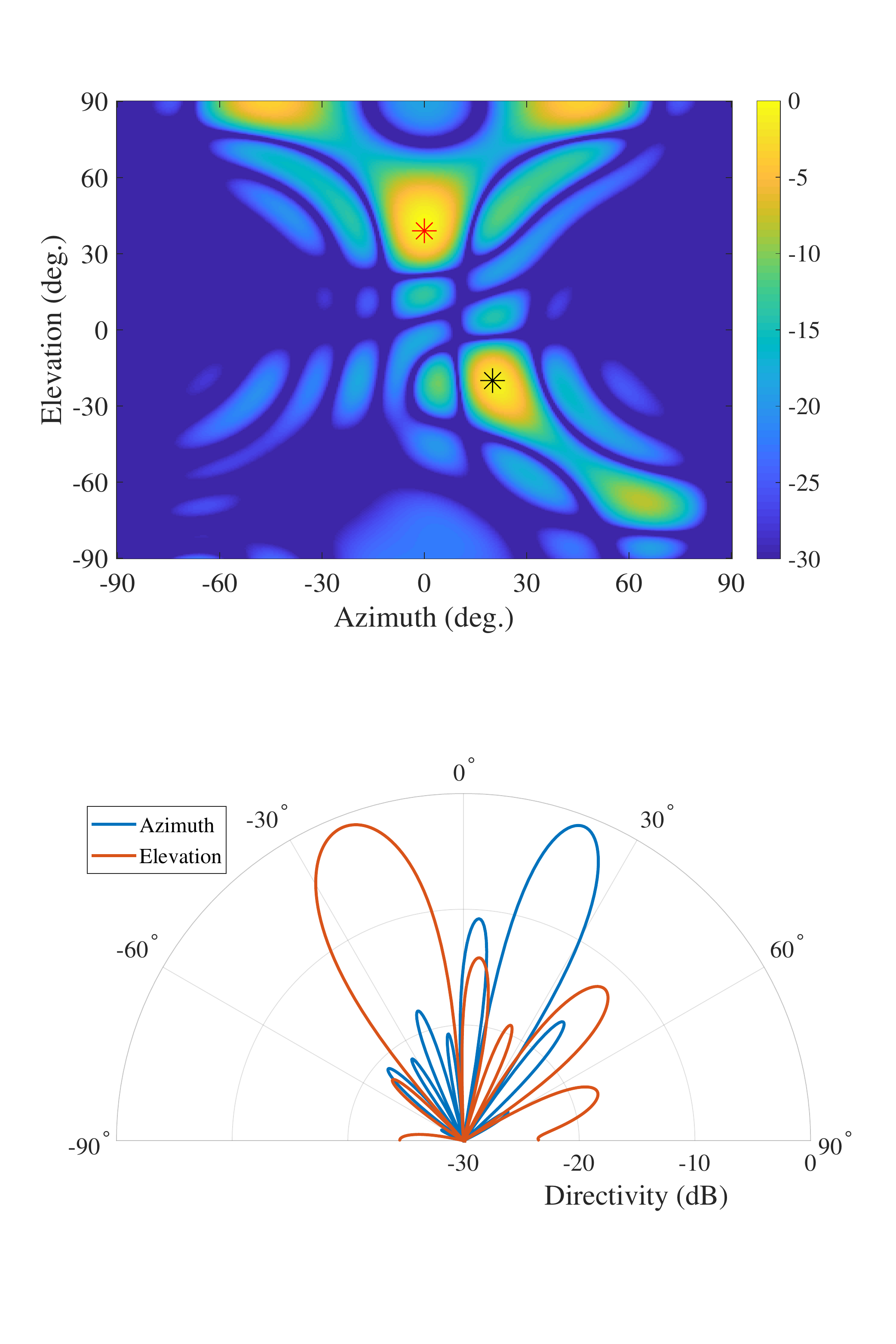}
	\caption{Normalized farfield pattern resulting from the simulated metasurface antenna array, forming a beam steered to $\theta=-20^\circ$, $\phi=20^\circ$ (the black star shows the target location while the red star shows the grating lobe).}
	\label{f:farfield_no_feed_steered}
\end{figure}

\noindent With the constrained Lorentzian scheme for choosing the polarizabilities, the radiation pattern consists of the main beam and potentially only one diffracted order.

Array factor calculations were used to model the beamforming performance of a metasurface antenna array shown in Fig.~\ref{f:diagram}, which comprises eight air-filled rectangular waveguides, each with twenty elements. Each waveguide has a width $a_r=2.0$~cm and is fed with the same phase. The antenna operates at 10~GHz and the elements within each waveguide are spaced at 0.5 cm ($\lambda/6$). Each element is analytically modeled, according to Eq.~\ref{eq:lorentzian}, with the resonant frequency tunable from 9.8 to 10.2~GHz. We remain agnostic to any specific element design and tuning mechanism and follow the analytic model described in Eq.~\ref{eq:lorentzian} and shown in Fig.~\ref{f:element}.

Using the array factor described in Eq.~\ref{eq:lcm_farfield}, a broadside beam was generated with the sample metasurface antenna. Fig.~\ref{f:mapping} shows the comparison between the ideal polarizability expressed in Eq.~\ref{eq:ideal_alpha} (shown with the gray dashed line) and the realized polarizability expressed in Eq.~\ref{eq:alpha_lcm} (shown with the blue line, with metamaterial locations at the red markers); these plots show the response of the elements in one of the 1D waveguide-fed metasurfaces in the antenna array shown in Fig.~\ref{f:diagram}. Note that to achieve the best approximation to the linear phase advance required for a steered beam, the magnitude ends up with a periodic modulation rather than the ideally flat profile. Fig.~\ref{f:weights_no_feed} shows the metamaterial elements' dipole moments, which are analogous to the antenna weights of traditional radiators. Fig.~\ref{f:farfield_no_feed} shows that a broadside beam is created, but a large grating lobe appears. The source of the grating lobe can be traced back to the oscillating magnitude profile in Fig.~\ref{f:mapping} and Fig.~\ref{f:weights_no_feed} and the resulting coarse effective element spacing in the y direction. The grating lobe remains present when a beam is generated which is steered to $\theta=-20^\circ$, $\phi=20^\circ$, as shown in Fig.~\ref{f:farfield_no_feed_steered}.

\section{Grating Lobe Suppression}
\label{s:feed}

\subsection{Grating Lobe Derivation}
To better illustrate the source of the grating lobes, we mathematically isolate the grating lobe term from the array factor. By examining Eq.~\ref{eq:array_factor}, the second term can be seen as the beamsteering term, in which the phase of the guided wave has been counteracted and the steering phase term has been applied. Meanwhile, the first term shows the grating lobe term, where the grating lobe exists along the $\phi=0$ direction. Substituting $\phi=0$ into the first term of Eq.~\ref{eq:array_factor} leads to

\begin{equation}
    \sum_{n=1}^{N}e^{-j y_n(\beta + k \sin\theta)}.
    \label{eq:grating_lobe_isolated}
\end{equation}

\noindent From this equation, the grating lobe will appear at $\theta_g=\arcsin{(\beta/k)}, \phi_g=0^{\circ}$. For the simulated metasurface, this equation predicts a grating lobe at $\theta_g=41^\circ$, which is consistent with Fig.~\ref{f:farfield_no_feed} and Fig.~\ref{f:farfield_no_feed_steered}. It should be noted that a sufficiently high dielectric may avoid this phenomenon, but this often invokes additional loss.

\subsection{Waveguide Feed Layer}
In this section, we propose varying the phase exciting each radiating waveguide in a metasurface antenna array as a means of suppressing the grating lobes. This proposal can be illustrated mathematically by updating Eq.~\ref{eq:h_field} to include an arbitrary phase term applied to each waveguide-fed metasurface.

\begin{figure}
	\centering
	\includegraphics[width=6.75cm]{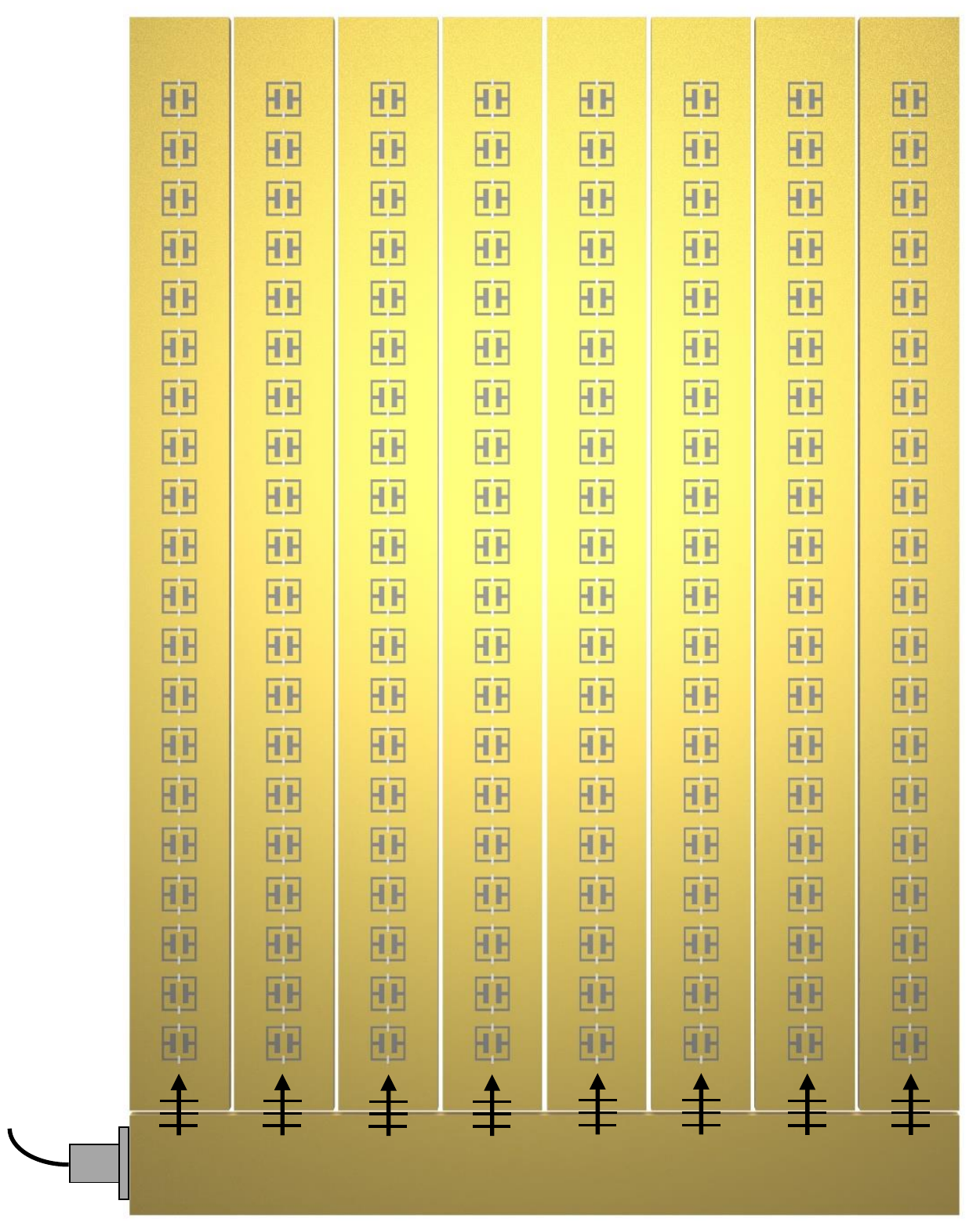}
	\caption{Diagram of a waveguide feed layer located at the bottom of the metasurface antenna shown in Fig.~\ref{f:diagram}. The waveguide feed layer simplifies the feed structure of the antenna to being a one port device and provides the phase diversity required to suppress grating lobes. Note that here, the waveguide feed layer is pictured on the same layer as the radiating waveguides, but the waveguide feed could also be located beneath the radiating waveguides.}
	\label{f:diagram2}
\end{figure}

\begin{equation}
\label{eq:h_phase_variation}
H_{n,m}=H_0e^{-j\beta y_n+j\gamma_m}
\end{equation}

\noindent Here $\gamma_m$ is the phase applied to the feed of the $m^{th}$ waveguide in the array. The optimal polarizability, as determined by Lorentzian-constrained modulation, then becomes

\begin{equation}
\label{eq:alpha_feed_adjusted}
\alpha_{n,m}=\frac{j-e^{j(\beta y_n +k x_m \sin\theta_s \sin\phi_s + k y_n \sin\theta_s\cos\phi_s-\gamma_m)}}{2}
\end{equation}

Combining Eq.~\ref{eq:array_factor} with Eq.~\ref{eq:alpha_feed_adjusted}, while incorporating Eq.~\ref{eq:h_phase_variation}, leads to a new array factor

\scriptsize
\begin{equation}
\label{eq:feed_array_factor}
\begin{split}
&AF(\theta,\phi)=\frac{H_0}{2}\Bigg[j\sum_{n=1}^{N}\sum_{m=1}^{M}e^{-jy_n(\beta  + k \sin\theta\cos\phi)}e^{-j (k x_m \sin\theta\sin\phi-\gamma_m)}\\
-&\sum_{n=1}^{N}\sum_{m=1}^{M}e^{-j k y_n (\sin\theta \cos\phi-\sin\theta_s \cos\phi_s)}e^{-j k x_m (\sin\theta \sin\phi-\sin\theta_s \sin\phi_s)}\Bigg]\\
\end{split}
\end{equation}
\normalsize

\noindent where $M$ is the number of waveguides. Eq.~\ref{eq:feed_array_factor} can be separated into two terms: the first term is the grating lobe term; the second term is the beamsteering term. The grating lobe term can be separated into the multiplication of two summations as

\begin{equation}
\label{eq:grating_lobe_term}
\frac{H_0j}{2}\sum_{n=1}^{N}e^{-j y_n(\beta  + k \sin\theta\cos\phi)}\sum_{m=1}^{M}e^{-j (k x_m \sin\theta\sin\phi-\gamma_m)}
\end{equation}

\noindent From Figs.~\ref{f:farfield_no_feed} and \ref{f:farfield_no_feed_steered}, the grating lobe from the metasurface behavior occurs along the $\theta$ direction, in the $\phi=0$ plane. To analyze the grating lobe more explicitly, we substitute $\phi=0$ into Eq.~\ref{eq:grating_lobe_term}. 

\begin{equation}
\label{eq:grating_lobe_term_reduced}
\frac{H_0j}{2}\sum_{n=1}^{N}e^{-j y_n(\beta + k \sin\theta)}\sum_{m=1}^{M}e^{j\gamma_m}
\end{equation}

\noindent In order to cancel the grating lobe term, the summation of $e^{j\gamma_m}$ from $m=1$ to $M$ must equal $0$. The most straightforward method for canceling the grating lobe then is select $\gamma_m$ as $\gamma_m=\pm m(2\pi/M)$, such that the term, $e^{j\gamma_m}$, is evenly spaced in the complex plane.

The optimal values for $\gamma_m$ could be realized with phase shifters (passive or active) or with a waveguide feed layer. Waveguide feed layers offer a small form factor, low loss, and have been used in previous works to excite antenna arrays \cite{hirokawa1992waveguide,wu2007design,li2014low}. If a feed layer were used to implement $\gamma_m$, it could be placed underneath the array and coupling irises could be etched at each radiating waveguide's location \cite{rengarajan1989characteristics,rengarajan1994accurate}. To suppress the grating lobes, the phase accumulation of the waveguide feed must match the desired $\gamma_m$ as

\begin{equation}
\label{eq:waveguide_feed_gamma}
e^{j\gamma_m}=e^{-jm\frac{2\pi}{M}}=e^{-j\beta_f x_m}.
\end{equation}

\noindent Here, $\beta_f$ is the propagation constant of the waveguide feed and $x_m$ is the position along the feed waveguide. When the feed layer samples the radiating waveguides at a spacing equal to the radiating waveguide width ($a_r$), this condition can be mathematically represented as $\beta_f a_r = 2\pi/M$, which is equivalent to

\begin{equation}
\label{eq:waveguide_match}
a_r\sqrt{\epsilon_0\mu_0\omega^2-\frac{\pi^2}{a_f^2}}=\frac{2\pi}{M}.
\end{equation}

\noindent Here, $a_f$ is the width of the waveguide feed. If this condition is satisfied, the waveguide feed layer will provide the optimal phase shift to each waveguide in the array and cancel the grating lobe. Rearranging the terms, Eq.~\ref{eq:waveguide_match} can be written as a design equation which determines the waveguide width of the feed ($a_f$).

\begin{equation}
\label{eq:waveguide_width_design}
a_f=\frac{M a_r \lambda}{2\sqrt{M^2a_r^2-\lambda^2}}
\end{equation}

\noindent Eq.~\ref{eq:waveguide_width_design} optimally suppresses the grating lobe, but can result in forcing the waveguide to operate close to cutoff if $M$ is too large. In such cases, substituting $M$ with an $M'$ which is a factor of $M$ (but greater than 1) will equivalently suppress grating lobes.

\begin{figure}
	\centering
	\includegraphics[width=8.8cm]{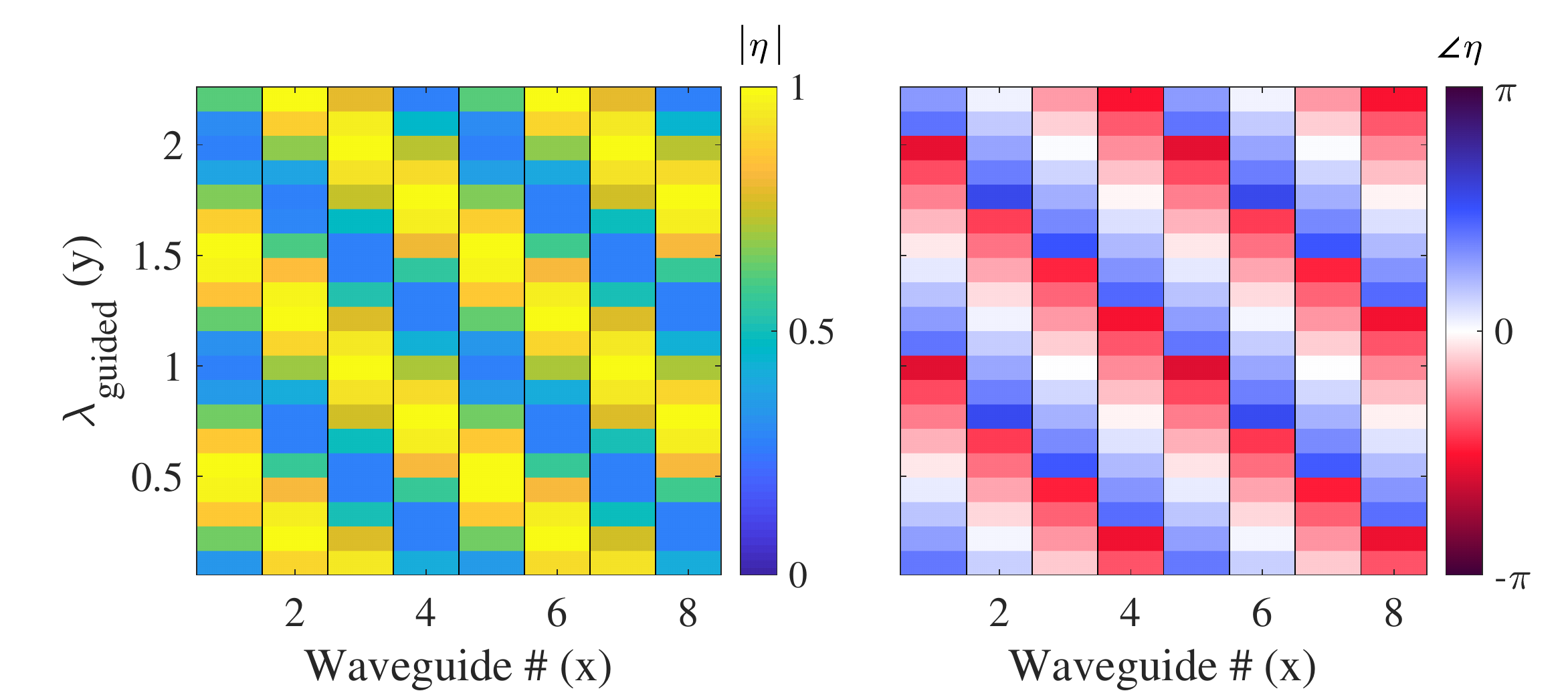}
	\caption{Antenna weights for the metasurface antenna with a feed layer.}
	\label{f:weights_with_feed}
\end{figure}

\begin{figure}
	\centering
	\includegraphics[width=7.5cm]{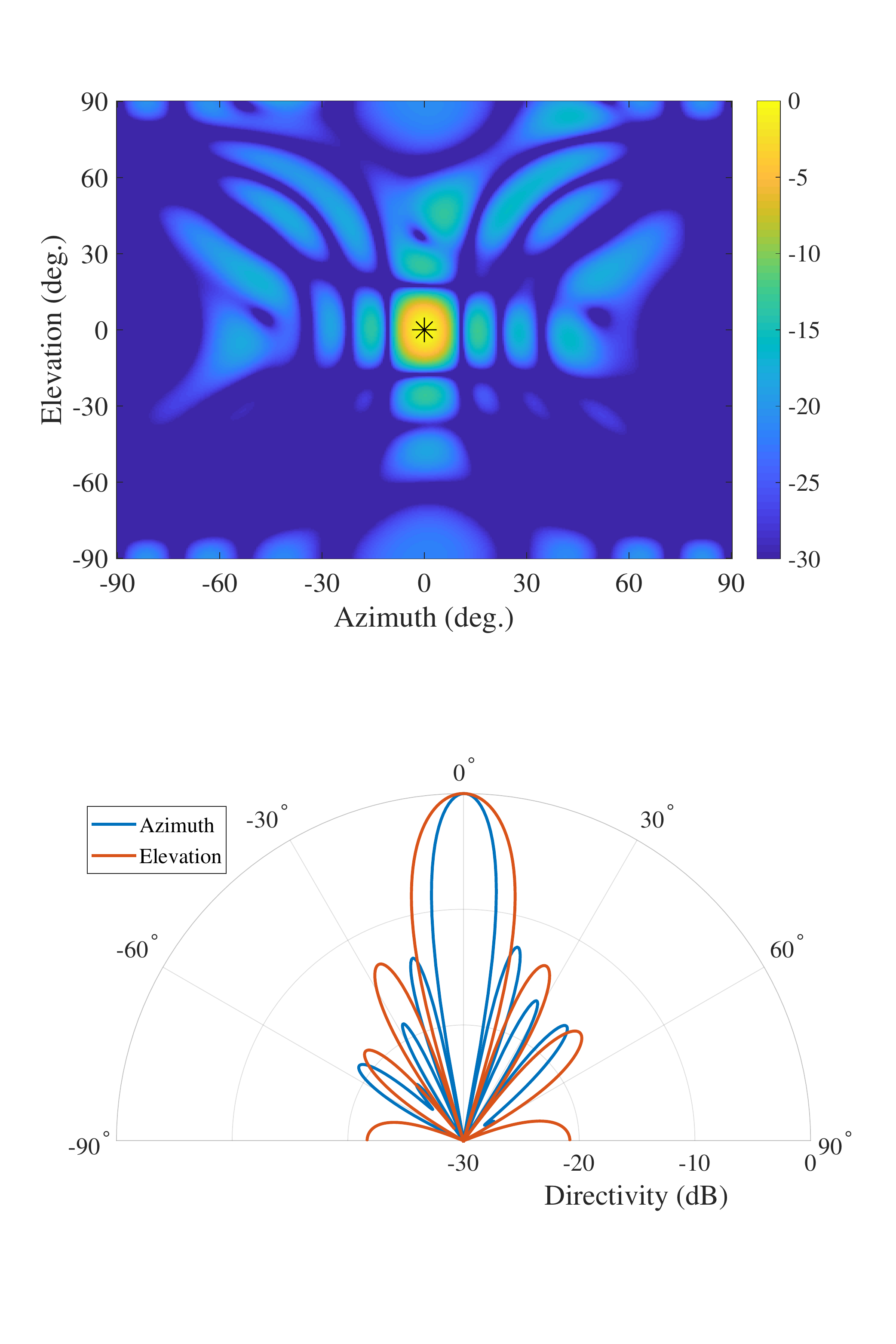}
	\caption{Normalized farfield pattern resulting from the simulated metasurface antenna array, which follows the mapping shown in Fig.~\ref{f:mapping} and \ref{f:weights_no_feed}. The beam is generated towards broadside, with the target location indicated by the black star.}
	\label{f:farfield_with_feed_broadside}
\end{figure}

\begin{figure}
	\centering
	\includegraphics[width=7.5cm]{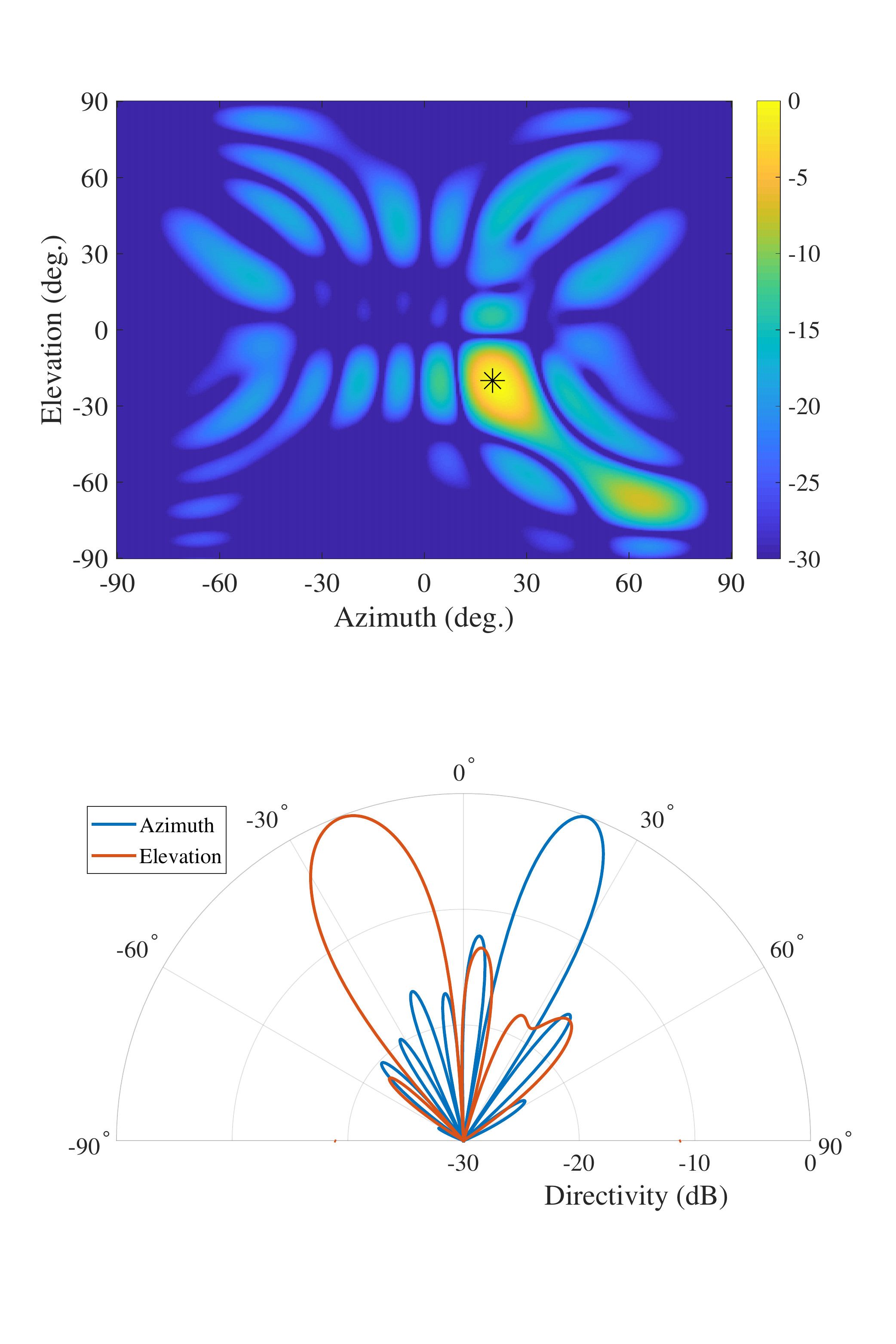}
	\caption{Normalized farfield pattern resulting from the simulated metasurface antenna array, which follows the mapping shown in Fig.~\ref{f:mapping} and \ref{f:weights_no_feed}. The beam is generated towards broadside, with the target location indicated by the black star.}
	\label{f:farfield_with_feed_steered}
\end{figure}

The result from Eq.~\ref{eq:waveguide_width_design} was used in an array factor calculation to create a feed layer to excite each waveguide in the antenna shown in Fig.~\ref{f:diagram}. In this particular metasurface antenna array, we use $M'=4$, since using $M=8$ would result in operation close to the cutoff frequency. Fig.~\ref{f:weights_with_feed} shows the updated antenna weights, which demonstrate the phase offset provided by the feed layer. The corresponding radiation pattern is shown in Fig.~\ref{f:farfield_with_feed_broadside} which clearly shows the grating lobe has been suppressed. Fig.~\ref{f:farfield_with_feed_steered} shows that when a beam is steered in both azimuth and elevation, the grating lobe remains suppressed.

\subsection{Frequency Dependence}
\begin{figure*}
	\centering
	\includegraphics[width=15cm]{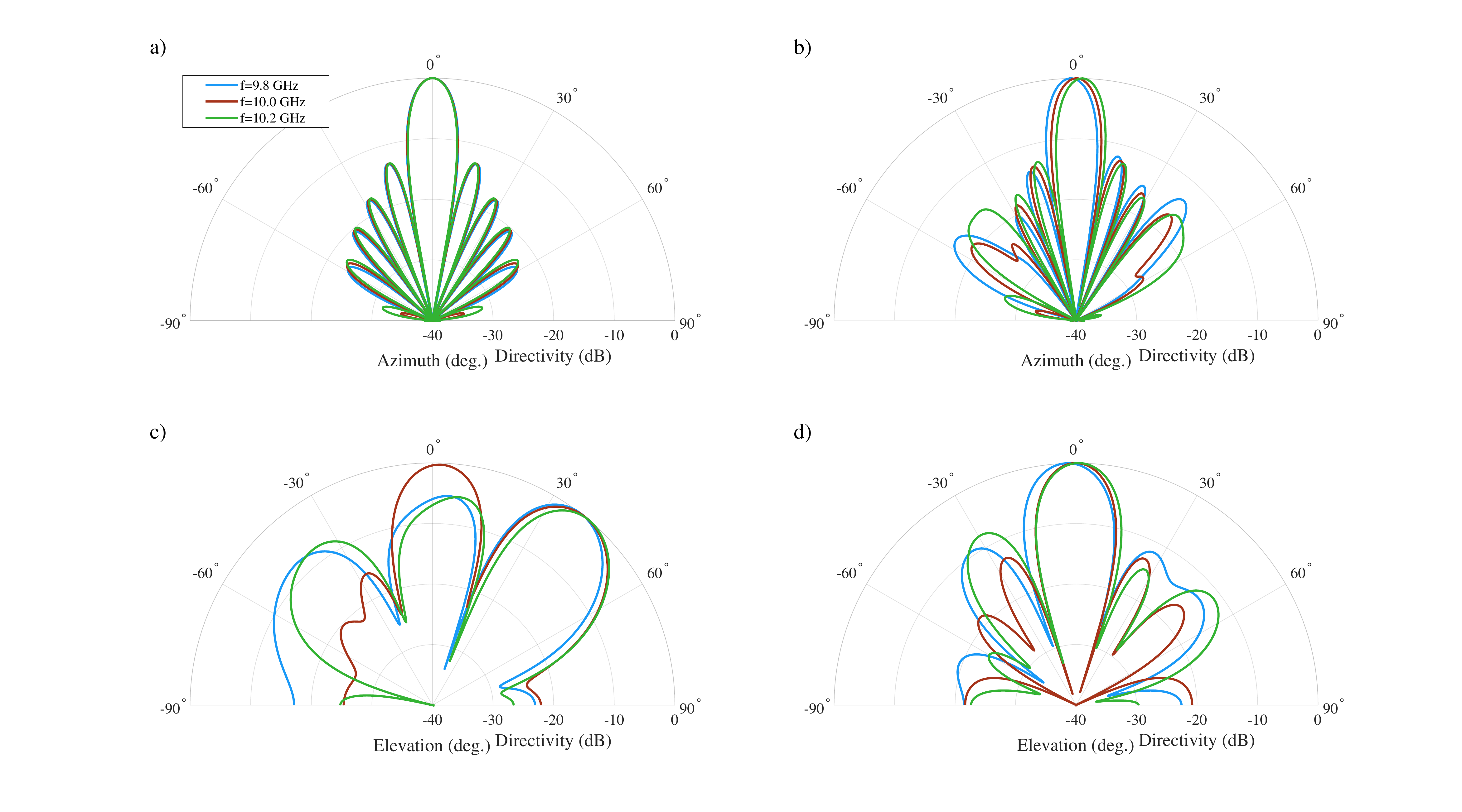}
	\caption{Normalized farfield patterns for 9.8, 10.0, and 10.2 GHz. a) shows the azimuth radiation pattern without the feed layer. b) shows the azimuth pattern with the feed layer. c) shows the elevation radiation pattern without the feed layer. d) shows the elevation pattern with the feed layer. In all cases, each element has been tuned for operation at 10.0 GHz.}
	\label{f:frequency_squinting}
\end{figure*}

To further analyze the effectiveness of the waveguide feed layer, we conduct a frequency study. We first examine the metasurface antenna's behavior when tuned to the center frequency and operating at other frequencies. For this study, we have generated a broadside beam with and without the feed layer, at 9.8, 10.0, and 10.2 GHz. From the elevation radiation patterns in Fig.~\ref{f:frequency_squinting}, the grating lobe is problematic at the operating frequency (10.0 GHz) and fully eclipses the steered beam when operating away from the target frequency. With the feed layer present, the grating lobe is well suppressed at the center frequency and remains mostly suppressed when operating away from the center frequency. It is also worth noting that when operating in this manner, the azimuth direction shows squinting behavior as a function of frequency, similar to a leaky wave antenna's operation, due to the existence of the feed layer.

\begin{figure*}
	\centering
	\includegraphics[width=15cm]{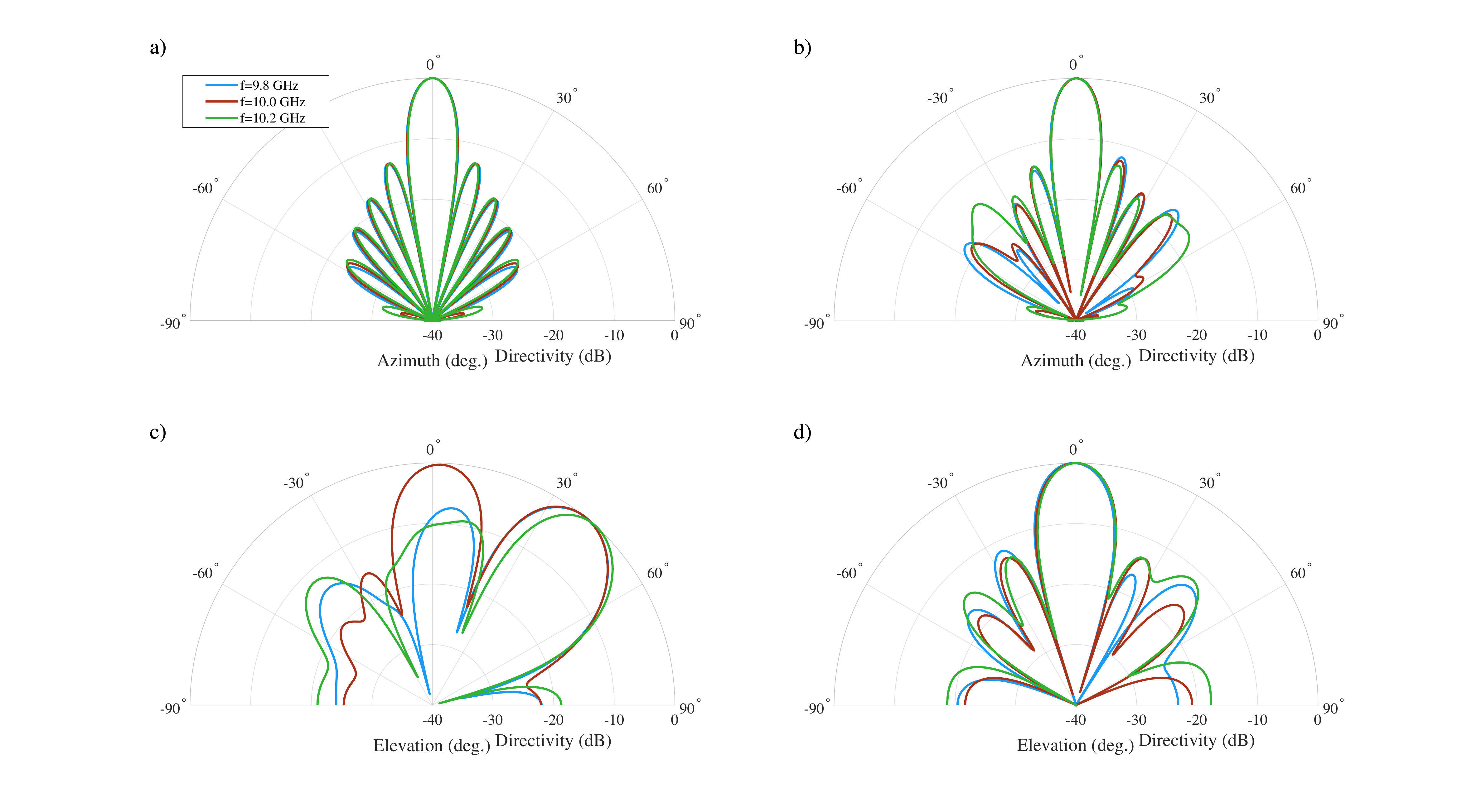}
	\caption{Normalized farfield patterns for 9.8, 10.0, and 10.2 GHz. a) shows the azimuth radiation pattern without the feed layer. b) shows the azimuth pattern with the feed layer. c) shows the elevation radiation pattern without the feed layer. d) shows the elevation pattern with the feed layer. For each operating frequency, a new tuning state has been optimized for each element.}
	\label{f:frequency_tuned}
\end{figure*}

Metasurface antennas can also consider another method of operation if integrated with components that enable high switching speeds. In addition to tuning the elements to the center frequency, it can be possible to update the tuning state of each metamaterial element as the frequency changes. In this context, a new tuning state is optimized and applied at each frequency. Here, we repeat the study above, but while updating the tuning state of each metamaterial element for each frequency. From the elevation radiation patterns in Fig.~\ref{f:frequency_tuned}, the grating lobe dominates the radiation pattern when the feed layer is not present. In addition, the azimuth frequency squinting seen in Fig.~\ref{f:frequency_squinting}b is avoided by re-optimizing the tuning state at each frequency. Further, tuning the antenna at each frequency shows improved performance, as seen by comparing the grating lobes in Fig.~\ref{f:frequency_squinting}d and Fig.~\ref{f:frequency_tuned}d.

\subsection{Different Tuning Strategies}

\begin{figure}[h]
	\centering
	\includegraphics[width=8.8cm]{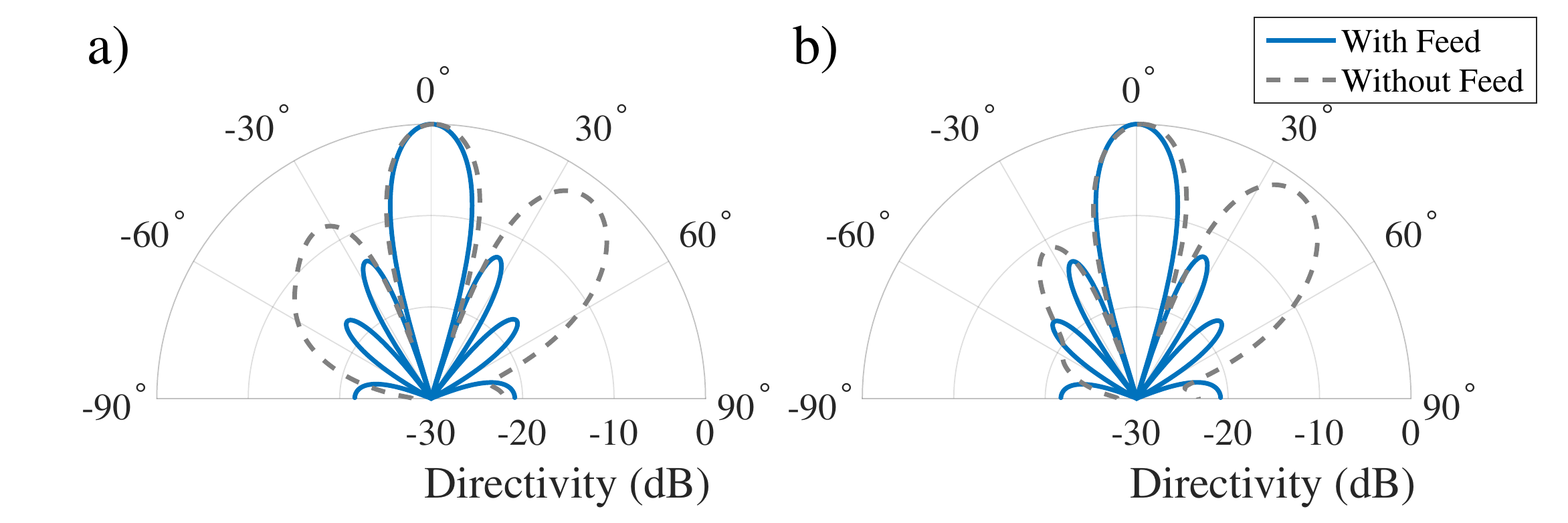}
	\caption{Normalized farfield pattern of the metasurface antenna array with and without a feed layer in the elevation direction. a) uses a direct phase matching approach to tuning the metamaterial elements. b) uses Euclidean modulation to tune the metamaterial elements.}
	\label{f:farfield_ph_em}
\end{figure}

We note that the need for a grating lobe suppression method is inherent to metamaterial elements, irrespective of the tuning strategy used. In addition to using Lorentzian-constrained modulation, we examined the same metasurface antenna with direct phase tuning and with Euclidean modulation. Both of these methods involve tuning the polarizability of each element to match the polarizability prescribed for beamforming in Eq.~\ref{eq:ideal_alpha}. In the case of direct phase tuning, the tuning state of each element is selected to minimize the phase difference between the polarizability from Eq.~\ref{eq:ideal_alpha} and the polarizability available as a function of tuning state. Euclidean modulation is similar, except that instead of minimizing the phase difference between these two quantities, the Euclidean norm between these quantities is minimized \cite{boyarsky2017alternative}. Euclidean modulation thereby considers the jointly coupled magnitude and phase response associated with metamaterial elements when tuning each element. Fig.~\ref{f:farfield_ph_em} shows radiation patterns from the simulated metasurface antenna using these methods. In both Fig.~\ref{f:farfield_ph_em}a and b, using the waveguide designed according to Eq.~\ref{eq:waveguide_width_design} suppresses the grating lobes, independent of tuning strategy.

\section{Conclusion}
Metasurface antennas stand to offer substantial hardware benefits as compared to existing electronically reconfigurable antennas, but the nature of metamaterial elements presents challenges with realizing the full potential of metasurface antennas. Without considering the differences in element behavior between traditional radiators and metamaterial elements, strong grating lobes can appear. A waveguide feed layer obviates the need for extremely dense element spacing or high dielectrics to avoid grating lobes and retains a metasurface antenna’s hardware benefits in terms of cost and manufacturability. The work in this paper outlines a roadmap for the implementation of large-scale metasurface antennas for applications requiring electrically-large reconfigurable antennas.

\newpage

\bibliographystyle{IEEEtran}

\bibliography{bib}

\end{document}